\documentclass[10pt, conference, compsocconf]{IEEEtran}
\usepackage{cite}

\ifCLASSINFOpdf
   \usepackage[pdftex]{graphicx}
   \DeclareGraphicsExtensions{.pdf,.jpeg,.png}
\else
   \usepackage[dvips]{graphicx}
   \DeclareGraphicsExtensions{.eps}
\fi
\usepackage{url}

\begin{document}
%
\title{Visualisation and Analysis Challenges for WALLABY}


\author{\IEEEauthorblockN{Christopher J. Fluke, David G. Barnes, and 
Amr H. Hassan}
\IEEEauthorblockA{Centre for Astrophysics \& Supercomputing\\
Swinburne University of Technology\\
Hawthorn Australia\\
Email: cfluke@swin.edu.au}
}


%


\maketitle

\begin{abstract}
Visualisation and analysis of terabyte-scale datacubes, as will
be produced with the Australian Square Kilometre Array Pathfinder (ASKAP),
will pose challenges for existing astronomy software and the work
practices of astronomers.   Focusing on the proposed outcomes of 
WALLABY (Widefield ASKAP L-Band Legacy All-Sky Blind Survey),
and using lessons learnt from 
HIPASS (H{\sc i} Parkes All Sky Survey), we identify issues 
that astronomers will face with WALLABY data cubes. We comment on 
potential research directions and possible solutions to these challenges.
\end{abstract}

\begin{IEEEkeywords}
computer aided analysis; distributed computing; radio astronomy; visualisation. 
\end{IEEEkeywords}

%
\IEEEpeerreviewmaketitle

\section{Introduction}

The Australian Square Kilometre Array Pathfinder (ASKAP; 
\cite{johnston:2007},\cite{westmeier:2010}),
represents a significant advance 
in radio telescope design.   This facility will combine high resolution
imaging (through the use of a 36-element aperture synthesis array 
with a maximum baseline of $6$~km) with a wide field of view (achieved 
with innovative focal plane array technology) at frequencies between 
700 MHz and 1.8 GHz.  Installation of
the first ASKAP antenna at the Murchison Radio Observatory site, Western
Australia, occured in early 2010, and the 6-antenna BETA test 
array will operate from September 2011-March 2013.  It is anticipated that 
full science operations will be underway by 2014. Processing and data transport 
requirements for ASKAP are described in \cite{cornwell:2008}, 
and \cite{quinn:2010} 
provides an overview of the 
data infrastructure requirements. 

WALLABY \cite{wallaby:url}
is one of ten ASKAP Survey Science Projects currently in the 
design study phase. WALLABY aims to significantly enhance our understanding of
the extragalactic neutral hydrogen (H{\sc i}) universe.  
The survey will cover 75\% of the
sky, detecting $\sim\!0.5$ million galaxies to redshift $z = 0.26$ (lookback time
$\sim\!3$ Gyr).
Key science outcomes are studies of galaxy formation and the missing 
satellite problem in the Local Group, evolution and star formation in galaxies, 
mergers and interactions, determination of the H{\sc i} mass function and its 
variation with galaxy density, and the nature of the cosmic web. 

Unlike previous H{\sc i} surveys, it will not be feasible to keep all of the raw 
data (i.e. Fourier visibilities) from ASKAP observations for subsequent 
reprocessing. Instead, pipeline-preprocessed spectral data cubes will 
be provided for analysis. Each WALLABY spectral cube is anticipated 
to comprise 6144 $\times$ 6144 spatial 
pixels and 16,384 spectral channels (i.e. $\sim\,600$ gigavoxels 
or volume elements in total),
requiring 2.5 terabytes (TB) of storage.
A total of 1200 cubes 
will be required to cover the sky south of declination $\delta = +30^\circ$.  
Likely additional outputs are integrated (moment) maps, continuum images, 
sub-cubes (individual objects or scaled versions of larger 
datasets), and full parameterisation of all galaxies, resulting in several
petabytes of data products.

Such data volumes pose considerable challenges for the existing work
practices of astronomers. 
Indeed, visualisation and
analysis (hereafter, ``V+A'') of WALLABY data products will require 
both evolutionary and revolutionary changes to existing software and 
hardware, with a likely move away from desktop-only solutions, 
and a greater reliance on remote services.   

A brief overview of the WALLABY workflow from data collection 
to catalogue is as follows:
\begin{enumerate}
\item Observe field.
\item Generate spectral data cube from visibilities.
\item Visualise cube as quality control prior to deletion of raw data.
\item Transfer preprocessed data cube to archive.
\item Perform source finding on data cube.
\item Fit models to candidates and perform related quantitative analysis.
\item Add parameterised candidates to catalogue.
\end{enumerate} 
Apart from personnel, the main resource for completion of all 
of these stages is access to appropriate computing infrastructure 
(hardware and software).  

As a framework within which to assess the practicalities of achieving 
each step in the WALLABY workflow, we begin (Section II) by considering 
desktop and high performance computing (HPC) cluster resources available 
and used by astronomers today, and project these forward to configurations 
available by 2014.  In Section III, we present five challenges that V+A of 
WALLABY data products will face in the likely computing 
environment.  We consider tasks that can be done essentially the same way 
they are now, and those requiring an investment in new technology or the
development of new software, in order to deal with data sets orders of magnitude
larger than previous extragalactic H{\sc i} survey projects.  
We make our concluding remarks in Section IV.

Throughout, we make comparisons with the existing H{\sc i} Parkes All Sky 
Survey (HIPASS; \cite{barnes:2001}), conducted 
with the Parkes Multibeam receiver \cite{staveley:1996}.
The southern catalogue, HICAT ($\delta < +2^{\circ}$; \cite{meyer:2004}), 
comprised 4315 galaxies, and the northern extension, 
NHICAT ($+2^{\circ} < \delta < +25^{\circ}30'$; \cite{wong:2006}), 
a further 1002 sources.  Russell Jurek (Australia Telescope National
Facility; ATNF) has combined the
388 individual southern sky data cubes into a single all-sky cube 
with $1721 \times 1721 \times 1025 =  3$ gigavoxels, 
and a file size of 12 GB.  

\section{The Compute Context}
The configurations of (typical) desktop and HPC resources available to 
astronomers are fundamental to the capacity of existing or new software 
to enable each stage of the WALLABY workflow.   
In Table I, we present hardware parameters for today's
mid-range desktop computer.  Using typical growth rates in the 
computing industry (e.g. Moore's Law; Kryder's Law \cite{walter:2005}), 
we extrapolate to 2014 (i.e. ``tomorrow'').  Quoted processing speeds 
are theoretical (i.e. peak), single precision values; these assume 100$\%$
efficient algorithms using all available processing cores/streams.
Table II presents similar per-node comparisons for cluster-based HPC configurations. 
Other HPC configurations are possible, but we restrict 
our discussion to facilities similar to the Swinburne ``Green Machine'' 
Supercomputer \cite{green:url}, with which we are most 
familiar.

Most specifications and capabilities of desktop and HPC compute 
platforms will simply evolve and grow as they have done over the 
past decades.  Two significant revolutions in compute capability of
processors, however, are currently underway (e.g. \cite{sutter:2005}):
\begin{enumerate}
\item Central procesing units (CPUs) 
are gaining increased compute capacity in the 
form of multiple cores, rather than increased clock speeds.
\item Graphics processing units (GPUs) are boosting compute capacity 
by around 10--50 times by 
functioning as streaming co-processors, at very low cost.
\end{enumerate}
This ``concurrency'' revolution, based on the availability of high 
levels of parallelism on a single chip, requires major software work 
\cite{kirk:2010}
and a re-examination of algorithms so that scientists can benefit fully 
from this new processing paradigm (see \cite{barsdell:2010}  and \cite{fluke:2010b}
for astronomy-related solutions). 
While CPUs are optimised for sequential programs thanks to sophisticated 
control logic and large memory caches (to reduce instruction and data
access latencies), GPUs maximise the chip area for computation.   The
advent of programming libraries such as CUDA and OpenCL has enabled the
use of GPUs for general purpose computation, with the GPU acting as a 
computational co-processor. Typical (single precision) theoretical peaks 
are already over 1 teraflop/s for cards like the NVIDIA Tesla C2050.
Some of the challenges we identify in this paper can only be solved 
with GPUs.


\begin{table}[!t]
\caption{Desktop computing for WALLABY}
\label{tbl:desktop}
\centering
\begin{tabular}{|l|l|l|}
\hline
{\bf Attribute} & {\bf Today (2010)} & {\bf Tomorrow (2014)} \\ 
\hline
CPU Clock & 3 GHz & 3 GHz \\
Number of Cores & 2 & 8 \\
CPU Speed & 24 gigaflop/s & 100 gigaflop/s \\
CPU Memory & 4 GB & 18 GB \\
CPU--Memory Bandwidth & 12 GB/s & 24 GB/s \\
CPU--Network Bandwidth & 10--100 MB/s & 100--1000 MB/s \\
Local Storage  & 0.5 TB & 5 TB \\
Local I/O Access & 50 MB/s & 100 MB/s \\ 
GPU Memory & -- & 2-3 GB \\
GPU Speed & -- & 2 Tflop/s  \\
GPU Memory Bandwidth &  -- & 200 GB/s \\
\hline
\end{tabular}
\end{table}

\begin{table}
\caption{Per-Node High Performance Computing for WALLABY}
\label{tbl:hpc}
\centering
\begin{tabular} {|l|l|l|}
\hline
{\bf Attribute} & {\bf Today (2010)} & {\bf Tomorrow (2014)}  \\
\hline
CPU Clock & 3 GHz & 3 GHz \\
Number of Cores & 8 & 24 \\
CPU Speed & 96 gigaflop/s & 500 gigaflop/s \\
CPU Memory & 16 GB & 72 GB \\
CPU--Memory Bandwidth & 12 GB/s & 36 GB/s \\
CPU--Network Bandwidth & 1--2 GB/s & 2-8 GB/s\\
Network Storage  & 10's TB & 1-10 PB \\
Number of nodes & 100-300 & 100-300 \\
GPU Memory & -- & 6-10 GB\\
GPU Speed & -- &  4-8 Tflop/s \\
GPU Memory Bandwidth & -- & 300 GB/s \\
\hline
\end{tabular}
\end{table}

%
%

\section{Challenges}
\subsection{Handling Big Data Files}
Steps 1--4 of the WALLABY work flow relate to producing spectral data cubes that
are significantly larger than have been available for previous surveys.
The logistics of moving large data cubes on the network notwithstanding, 
it should be immediately apparent that an entire WALLABY cube cannot 
fit in the main memory of either today's or tomorrow's desktop configuration, 
and only one full resolution cube (at a time) can be stored on tomorrow's 
internal desktop hard drive.  Indeed, 16 GB memory limits sub-cubes 
to 2k $\times$ 2k $\times$ 1k (= 4 gigavoxels) for in-memory analysis.  Since most existing 
astronomy software for the V+A of radio telescope data
(e.g. Karma \cite{karma:url}, AIPS \cite{AIPS:url}, CASA \cite{CASA:url}) is aimed
at handling data sets that can fit in the host memory of a desktop machine,
without further development, these packages are clearly incompatible 
with handling 2.5 TB cubes.

For V+A tasks that require access to an entire spectral cube 
(see below), the practical alternative is to use a distributed computing 
cluster architecture as a remote V+A service.  This is one
of the anticipated roles of the Pawsey HPC Centre \cite{quinn:2010}, 
however, other major
computing facilities such as the Swinburne Supercomputer incorporating 
gSTAR (the GPU Supercomputer for Theoretical Astrophysics Research)
could also be used.
In principle, 2.5 TB of memory must be available across a
computing cluster: assuming 16 GB (or 72 GB) is available per compute node, 
this means 160 machines today (or 36 tomorrow); there are clear advantages 
in managing fewer machines, each with more memory. 

Moving software 
to a cluster environment neccesitates the use of a distributed 
memory infrastructure, and an understanding of the level of parallelism 
in V+A algorithms.  A data-parallel paradigm will be appropriate 
in many cases.

A remote service mode of operation is not likely to have a negative impact on
the user's experience for most large-scale analysis tasks 
(e.g. source finding or  re-gridding), as these do not occur 
in ``real-time''.  The ability to achieve interactive visualisation at 
frame rates 
above 5-10 frames/sec (fps) will be limited by factors such as processing 
and network speed, and bandwidth.

To maximise efficiency, a distributed cluster also requires a parallel 
file system or other form of distributed network storage.  Unfortunately, 
astronomy's standard FITS (Flexible Image Transport 
System \cite{FITS:url})
file format is not ideal for 
parallel access. Practical alternatives for faster access include 
NetCDF \cite{NetCDF:url}
or HDF5 \cite{HDF5:url} formats,
but these require either ``on-the-fly'' transformations 
between file formats and metadata, or a possible need to increase 
the total storage for the WALLABY survey cubes.

\subsection{Global Views versus Image Slices}

The need to discard raw visiblity data from ASKAP early in the WALLABY 
workflow (Step 3) means that global quality control of data cubes 
will be critical.  
Possible noise characteristics and artefacts may include large-scale
gradients, non-uniform noise levels across the field of view, incompletely 
subtracted continuum sources and hydrogen recombination lines.

While inspecting individual slices may be one approach to quality control, 
this is not straightforward.  Suppose it was possible to sequentially
examine individual 2D slices from a WALLABY data cube 
(along the spectral axis), at a reaonsable frame inspection rate of 5 fps: 
it would take $\sim\!1$ hour to step through 16k spectral slices.  This 
assumes a display capable of displaying 6k x 6k pixels - for a 
HD-1080 monitor, we require at least 3x6 sub-cubes, thus increasing
the view time to 18 hours per cube. Moving to sub-cubes may limit opportunities
to understand global variations.  Moreover, slicing techniques remove
the ability to perceive artefacts or noise characteristics along the slicing 
axis, so it may be necessary to slice along more than one axis. 
Alternatively, scaled down cubes could be inspected, but these may hide 
artefacts, as scaling of approximately 10:1 (spatially) would be required. 


A preferred option may be to use a multi-panel display for full resolution 
visualisation.  For example, the 
OptiPortal \cite{Opti:url}
at the Commonwealth Scientific and Industrial Research Organisation 
(CSIRO) ICT Centre, Marsfield, New South Wales, comprises 25 
high definition panels, with a total resolution over 50 
megapixels. Accordingly, a full resolution WALLABY cube could be viewed on 
such a display at a 1:1 mapping of data to screen pixels.  

A GPU-cluster based visualisation framework capable of volume rendering ``larger
than memory'' data cubes at interactive frame rates has been demonstrated
by \cite{hassan:2011}.  In recent tests of
this framework using the CSIRO GPU Cluster in Canberra (256 Tesla S2050 GPUs with
3 GB/GPU), frame rates of better than 40 fps (1024 x 1024 pixel output) were 
achieved
for a 204 GB cube  using 128 S2050 GPUs. 
Scaling this to a full WALLABY data cube 
requires a minimum of $\sim\!450$ x 6 GB GPUs (or $\sim\!275$ x 10 GB) 
in 2014.  This task will not be feasible at interactive frame rates with a 
CPU-only HPC cluster.
A combination of an OptiPortal and a GPU cluster may support fully
three-dimensional global views of 
WALLABY data cubes, and the ability to quickly identify (compared with slicing) 
areas of a data cube that may indicate further processing is required using 
the visibilities.  


\subsection{Source Finding and Confirmation}
Source finding is the process of identifying and extracting candidate sources from
a data cube.  To a large extent, the science outcomes of WALLABY depend on the existence
of source finding software that maximizes reliability (i.e. {\em only} identifies 
extragalactic H{\sc i} sources) and completeness (i.e. finds {\em every} source that 
exists within a data cube).  An ideal source finder would have 
a 1:1 candidate to source ratio, and offer 100\% completeness.  

Conceptually, source finding is a simple task: examine each voxel in turn and determine 
the amount of source signal contributing to that voxel.  Practically, source finding 
is extremely difficult, as every voxel contains both source and non-source components.
The latter include noise (that may vary across the field of view), interference 
(natural and artificial), contamination from bright sources outside the field, 
recombination lines, incomplete subtraction of continuum sources, and so on.  

It is instructive to consider the source finding tasks undertaken for HIPASS.
The southern HIPASS catalogue, HICAT, used two main source finders: 
{\sc multifind} and {\sc tophat}. These produced $\sim\!140,000$ candidates, 
all of which were inspected manually (see \cite{meyer:2004} for details of these
source finders). Neither source finder identified all candidates in the
final source list.  The overall performance of {\sc tophat} was much better than 
{\sc multifind}: 17,232 {\sc tophat} candidates resulted in $90 \%$ 
of the final catalogue of 4315 galaxies.  Due to its lower candidate-to-source
ratio, only {\sc tophat} was used for the northern extension, NHICAT, 
with 14,879 candidates resulting in 1002 astronomer-confirmed 
sources \cite{wong:2006}.

For HIPASS, it was possible to view $>150,000$ candidates by eye
in order to provide confirmation of source identification. Overall,  
$\sim\!95\%$ of candidates were rejected.  Limiting this to {\sc tophat}, 
the rejection rate was $75\%$ for HICAT and $93\%$ for NHICAT. For the expected
$\sim\!0.5$ million WALLABY sources, such high rejection rates will be crippling if
human inspection is expected to play a significant role.  Assuming a perfect
source finder (i.e. no false detections) and 1 minute per source to load data, 
confirm, and annotate a candidate for later analysis, inspecting 0.5 million
candidates will take a minimum of $\sim\!1$ year (walltime). Fortunately, 
this is a parallel task that can commence before all survey cubes 
are obtained. The inspection processes could be shared between WALLABY
team members, provided consistency in source confirmation can be assured.

As with visualisation, source finding within 2.5 TB data cubes requires (at minimum) a
distributed computing approach.  Effort is underway to produce a distributed
version of {\sc duchamp} \cite{DUCHAMP:url}, but as with 
HIPASS, more than one source finder may be required.  While solutions to the source 
finding problem are outside the scope of this paper, we assert that a GPU cluster 
will prove to be beneficial here. For compute intensive tasks, GPUs offer
a massive processing gain at much lower cost than a CPU-only cluster with the
equivalent processing power. GPUs may also permit alternative approaches 
to source finding that are simply not feasible to undertake on a CPU. 




\subsection{Desktop Visualisation and Analysis}
Assuming we have solved the data handling problem, and that an appropriate catalogue
of sources is available for inspection and quantitative analysis, we now consider 
what could be achieved on a desktop computer in 2014.  

The biggest limitation is likely to be the amount of main memory: 16 GB will 
accommodate a $\sim\,$4 gigavoxel cube, with a choice between cropping 
and subsampling from a larger WALLABY cube.  Storing the WALLABY data in 
16 GB tiles would facilitate some reasonable level of ``traditional'' 
handling of data by astronomers - but the entire survey would now occupy 
nearly 190,000 tiles instead of 1200 cubes.

A reasonable balance between the spatial and spectral 
axes yields:
\begin{equation}
\left( \frac{s_{\alpha\delta} d_{\alpha\delta}}{6} \right)^2 
\left( \frac{s_z d_z}{4} \right) > 1
\end{equation}
where $s_{\alpha\delta}$ ($s_z$) is how coarsely the user is prepared to 
subsample along each spatial (spectral) axis of the WALLABY data cube, 
and $d_{\alpha\delta}$ ($d_z$) is the factor by which the user is prepared to 
crop a standard WALLABY cube along each spatial (spectral) axis.

Let us refer to a cube that has been subsampled and/or cropped to fit 
in main memory on tomorrow's desktop as a {\em scube}.  Scubes will be acceptable 
for most modes of qualitative visualisation, but are not appropriate 
if quantitative analysis 
is going to be attempted; here, cropping is the necessary choice, but this will 
substantially reduce the area of sky and/or frequency space that is represented 
by a single scube.  

While a 16 GB scube can easily be stored locally, it will take nearly 3 minutes 
to load into memory - waiting for data to load will become a much more common task 
for tomorrow's astronomers.  Once loaded, even the simplest of traditional 
operations (e.g. find the minimum, maximum, mean, standard deviation) will take on 
the order of seconds in the absolute best case (based on having to process 
the entire scube through the CPU).
If any significant additional processing or filtering of the scube is desired,
then the desktop platform will not have sufficient compute capability in the CPU 
alone.  
A desktop platform with a GPU co-processor would improve the situation, 
but not drastically, as the scube is still too large to fit on the GPU’s own 
local memory (2-3 GB).  The improvement in compute capability might in practice be a 
few times, but is unlikely to be better than 10 times.

Assuming a 4 gigavoxel WALLABY scube, the following traditional, 
interactive visualisation tasks should be feasible on tomorrow's desktop:
\begin{enumerate}
\item Image slicing: 4 gigavoxels can be scanned on a 1 megapixel display, 
at 25 fps, in under three minutes.  Compare this with HIPASS: 1024 frames 
at 25 fps takes $\sim\,$40 seconds, but there is much less data 
(HIPASS cubes had spatial 170 x 160 pixels, with 
some blanked, so the information content is vastly lower per frame).
\item Volume rendering: to accomplish a traditional, hardware-accelerated 
texture-based volume rendering, we must further compress our scube from 
16 GB down to $\sim 2$ GB (500 megavoxel $\sim\!800^3$ voxels) so that the image 
fits in the GPU co-processor memory.  Subsampling is likely the preferred option here, 
as traditional volume rendering is qualitative not quantitative.  
\end{enumerate}

\begin{table}[!t]
\renewcommand{\arraystretch}{1.3}
\caption{Three-dimensional texture volume rendering performance
using NVIDA GT120 GPU with 512 MB RAM. }
\label{tbl:volrender}
\centering
\begin{tabular}{|c|c|c|c|c|}
\hline
&& {\bf Minimum} & {\bf Maximum} & {\bf Average}\\
${\mathbf N}$ & ${\mathbf N^3}$ {\bf voxels} &
 {\bf (fps)} & {\bf (fps)} & {\bf (fps)} \\
\hline
150 & $3.4 \times 10^6$ & 20.0 & 20.0 & 20.0 \\
200 & $8.0 \times 10^6$ & 10.0 & 19.9 & 14.0 \\
250 & $15.6 \times 10^6$ & 8.6 & 11.2 & 15.0 \\
300 & $27.0 \times 10^6$ &  6.7 & 12.0 & 9.1 \\
350 & $42.9 \times 10^6$ &  6.0 & 10.0 & 7.8 \\
\hline
\end{tabular}
\end{table}

While fitting a 2 GB scube into GPU memory is achievable, we still require
an interactive frame rate of $>5$ fps.  Table III presents the results
of performance testing with an NVIDIA GT120 GPU (512 MB RAM).  
Today's desktop with a mid-range GPU can render up to $350^3$ voxels, 
filling 600 x 600 pixels on the screen, at $\sim\,$8 fps.  This limit 
is imposed by the maximum texture size on the card (a factor of 
both the hardware and the application programming interface).  
A top-end graphics card 
today (the ATI Radeon 5970) can render a $500^3$ volume 
(125 megavoxels = 500 MB) at around 8 fps comfortably, just filling 
$1000 \times 1000$ pixels when the cube is face on. 
In practice this three-dimensional (3D) texture rendering does better for certain                     
orientations of the cube, presumably corresponding to more continguous            
memory access when gathering textures from the volume.  Thus 8 fps is a 
conservative lower bound; around half the time it is actually 
managing closer to 15 fps.   

We can extrapolate our results to estimate a rendering rate 
of $\sim\,$1 fps if we could fit a 500 megavoxel cube on the card.  On tomorrow's 
desktop platform this could be accomplished at $\sim\,$4 fps.  Not a stellar 
result, so even tomorrow, texture-based volume rendering will be limited by 
rendering capability, not GPU memory size.


%
%
%


Both the Local Volume H{\sc i} Survey (LVHIS; \cite{koribalski:2008}) and The
H{\sc i} Nearby Galaxy Survey (THINGS; \cite{walter:2008}) have 
demonstrated the diversity 
in H{\sc i} kinematic structures in the local universe.  Simple models, such
as differentially rotating H{\sc i} disks \cite{rogstad:1974}, \cite{begeman:1989},
do not capture the complexities of warps, anomalous gas and mergers. A 
typical modelling process involves the generation of six-dimensional 
position and velocity values for an input model, and mapping these to 
two spatial coordinates and a line-of-sight velocity. 
New opportunites may arise for visualisation-directed, interactive 
model-fitting to complex kinematic structures using an approach of the
type described by \cite{fluke:2010a}.  The highly data parallel nature of this 
processes (the contribution of each spatial pixel, and hence line-of-sight,
can be computed independently of all others) is well-matched to the 
GPU, so interactive frame rates are unlikely to be computationally limited. 
See also the {\sc shape} 3D modelling tool for a similar technique applied to 
planetary nebulae and other bipolar outflows \cite{steffen:2010}.

\subsection{Data Product Management}
While overall ASKAP data management will be largely the responsibility of the
Pawsey HPC Centre, individual survey projects will need to 
carefully consider how they approach management of derived data. 
For a survey as comprehensive and data-rich as WALLABY, there is no place
for the somewhat ad hoc data management practices that have sufficed for
earlier all-sky extragalactic H{\sc i} surveys. 
The access times required to open and edit files notwithstanding, text
files are not a satisfactory solution for managing catalogues of 
$\sim\,0.5$ million galaxies, plus similar orders of rejected or 
unconfirmed candidates.

Catalogues will need to capture model parameters, reasons for rejecting
candidates,  meta-data relating to the provenance of analysed sources
(which analysis package was used, by whom, and with what set of input
parameters, so that the results can be repeated). 
Moreover, it will necessary to share up-to-date modifications of the
catalogue between multiple collaborators.
Solutions here are likely
to include large-scale commercial databases - and may be one of the cases
where astronomers should spend money to buy a solution, rather than 
reinvent one themselves.  We intend to address data management solutions
for WALLABY in future work, but note that understanding the benefits and 
limitations of approaches used for similar large-scale catalogues
from observational (e.g. 
Sloan Digital Sky Survey \cite{SDSS:url} and WiggleZ \cite{WiggleZ:url})
and simulation (e.g. Millenium \cite{Mill:url})
projects will be essential.

\begin{table}[!t]
\caption{Visualisation and analysis challenges for WALLABY, with potential
solutions. }
\label{tbl:vischal}
\centering
\begin{tabular}{|l|l|c|}
\hline
{\bf Challenge} & {\bf Solution} \\  \hline
Big data files & Use distributed file system \\
& and remote V+A services. \\
Global views & Use cluster of GPUs and \\
& large-format  displays. \\
Source finding & Requires most attention. \\
& GPU-cluster approach beneficial?\\
Human inspection & Not feasible without \\
& high-reliability source finders. \\
Desktop visualisation & Use GPUs for computation  \\
& and display.  \\
Image slicing & Only practical for sub-cubes. \\
Quantitative analysis & Opportunities for automated \\
& and interactive fitting with GPUs. \\
Data management & Must not be ad hoc. Databases \\ 
& must be used wisely.\\
\hline
\end{tabular}
\end{table}

\section{Concluding Remarks}
Perhaps the biggest challenge to planning strategies for visualisation
and analysis is that no ASKAP data exists yet. We do not know what the exact imaging 
properties of ASKAP will be. Although simulated data cubes are now being
generated, until the full ASKAP system undergoes commissioning, we will not fully
understand all of the calibration, noise, interference, etc. issues that will 
arise with the relatively new technology of focal plane arrays.  

Testing source finders often includes injecting fake sources, with a given 
signal level, and then seeing how often they are recovered. With real WALLABY 
data cubes unavailable until 2014, progress in testing source finders will
necessarily be limited.  
While we can 
do our best to plan source finders based on existing datasets, and early science
data from the BETA configuration (September 2011-March 2013), we may find that
our techniques do not work adequately for the full dataset.  By considering 
the various V+A tasks now, and identifying approaches based on new hardware
and software that were not available or feasible for earlier surveys, we can
hope to minimise the impact of the ``unknown unknowns'' of ASKAP.

Graphics processing units offer an intriguing solution to a number of the
current desktop-bound problems.   Table IV summarises our thoughts regarding the
visualisation and analysis tasks that will require either an evolution of
existing software and hardware, or a revolution in how they are approached.
By planning today, we aim to maximise the scientific return from WALLABY 
tomorrow.

\section*{Acknowledgment}
The authors would like to thank B\"{a}rbel Koribalski (ATNF) and 
Russell Jurek (ATNF) for discussions relating to the WALLABY project, and
John Taylor (CSIRO) and Tim Cornwell (ATNF) for enabling access to the
CSIRO GPU cluster. 



\bibliographystyle{IEEEtran}
%

\end{document}